\documentclass{aa}
\usepackage{graphics}

\begin{document}
  \thesaurus{09        
stars
              ( 09.16.1;  
                09.01.1;  
                08.23.2)}  
\title{Galactic planetary nebulae with Wolf-Rayet nuclei.\\ II. A 
consistent observational data set}
\fnmsep\thanks{Based on data obtained at the
Observatorio Astron\'omico Nacional, SPM, B.C., M\'exico}
\author{M. Pe\~na\inst{1},  G. Stasi\'nska\inst{2} \and S. Medina\inst{1}}

\institute{Instituto de Astronom{\'\i}a, Universidad Nacional
Aut\'onoma  de  M\'exico, Apdo. Postal 70 264, M\'exico D.F. 04510,
M\'exico\\ e-mail: miriam@astroscu.unam.mx; selene@astroscu.unam.mx
  \and DAEC, Observatoire de Paris-Meudon, 92195 Meudon Cedex,
France\\e-mail:
grazyna.stasinska@obspm.fr
  }
\date{Received ~~~~~~~~~~~~~/ Accepted  }
\titlerunning{WR Planetary Nebulae in the Galaxy}
\authorrunning{Pe\~na, Stasi\'nska \and Medina}
\offprints{M. Pe\~na}
\maketitle
\begin{abstract}
We present high resolution spectrophotometric data for a sample of 34
  planetary nebulae with [WC] spectral type central stars (WRPNe) in
  our Galaxy.
   The observed objects cover a wide range in stellar
characteristics: early and late [WC] type stars, as well as weak-emission
line stars (WELS).
Physical conditions in the nebulae (electron density and temperatures)
  have been obtained
from various diagnostic line ratios, and chemical abundances have
been derived with the usual empirical scheme. Expansion velocities
were estimated in a consistent manner from the line profiles for most
  objects of the sample.
A statistical study was developed for the derived data in order to find
fundamental relationships casting some light on the evolutionary status
of WRPNe.

We found evidence for a strong electron temperature gradient in WRPNe
which is related to nebular excitation. Such a gradient is not
predicted in simple photoionization models.

Abundance ratios indicate that there seems to be no preferential stellar mass
for the Wolf-Rayet phenomenon to occur in the nucleus of a planetary nebula.

Two objects, M\,1-25 and M\,1-32, were found to have a very small Ne/O ratio,
a property difficult to understand.

We reexamined the relation between
the nebular properties of the WRPNe and the spectral types of the central
stars. Our data confirm the trend found by other authors of the 
electron density
decreasing with decreasing spectral type, which was interpreted as
evidence that [WC] stars evolve from late to early [WC] types.
On the other hand, our data on the expansion velocities do not show the 
increase of expansion velocity with decreasing spectral type, that one might
expect in such a scenario.

Two objects with very late [WC] type central stars,  K\,2-16 and PM\,1-188,
do not follow the general density sequence, 
being of very low density for their spectral types. 
We suggest that the stars either underwent a late helium flash 
(the ``born again'' scenario)
 or that they have had a particularly slow evolution from the AGB. 

The 6 WELS of our sample follow the same  density vs. [WC]-type relation
as the bona fide WRPNe, but they tend to have smaller expansion
velocities. Considerations about the evolutionary status of WELS must
await the constitution of a larger observational sample.

The analysis of the differences between the WRPNe in the Magellanic
Clouds (distribution of [WC] spectral types, N/O ratios)
and in the Galaxy indicates that metallicity affects the [WR] phenomenon
in central stars of planetary nebulae.

  .
\end{abstract}
\section{Introduction}
Since the pioneering work of Paczy\'nski (1971), the basic scheme for the
production of planetary nebulae (hereinafter PNe) and the evolution 
of their nuclei  is
relatively well understood. Planetary nebulae are formed from stars of
initial masses between 1 and 8 M$_\odot$, which have completely burnt 
hydrogen and
helium in their cores, have twice  climbed the giant branch and lost
most of their mass through intense winds. At this stage, the stars
consist of dense
carbon-oxygen cores of about 0.6 M$_\odot$ surrounded by an
envelope composed of two thin shells.
The innermost one is composed of helium, and the outermost
of hydrogen. When the strong winds cease, the stars evolve towards
larger effective temperatures and start ionizing the matter lost 
previously in the winds.
The physical  reason for the departure from the asymptotic giant branch is
not quite understood,
nor is the nature of the energy source (hydrogen- or helium-burning) 
during the evolution of the
planetary nebula nuclei (Iben 1995).

About 50 planetary nebulae in our Galaxy are known to have central stars
of Wolf-Rayet (hereinafter [WR]) type among 350 PNe with a stellar
continuum measured and over 1000 PNe with nebular spectroscopy
available (Tylenda et al.  1993; G\'orny \& Stasi\'nska 1995).
All these [WR] central stars have been reported to be of [WC] spectral type,
mostly [WC\,2--4]  and [WC\,8--11]
(Tylenda et al. 1993) and their atmospheres are almost pure helium and
carbon (e.g. Hamann 1997). These planetary nebulae with [WR] nuclei,
hereinafter WRPNe, have recently been the subject of much attention
(e.g. G\'orny \& Stasi\'nska 1995; Crowther et al. 1998;  G\'orny \&
Tylenda 2000; G\'orny et al. 2000),
as they pose a problem for the theory of the evolution of
PN central stars. Indeed, helium-burning
models  (Wood \& Faulkner 1986; Vassiliadis \& Wood 1994) do maintain
a thin  hydrogen-rich outer layer, as emphasized by G\'orny \&
Tylenda (2000). One way to obtain a hydrogen-free outer layer is
through the born-again scenario (Iben et al. 1983), when a final
helium-shell flash occurs while the star is in its cooling phase, and
drives it back to the asymptotic giant branch (AGB). However, such a
scenario cannot hold for all the WRPNe, as shown by G\'orny \&
Tylenda (2000). They suggest that the central stars of most WRPNe
evolve directly from the AGB. The high carbon abundance observed in
these stars implies that deep mixing has occurred (Herwig et al.
1997; Herwig et al. 1998).

It has been suggested that the spectral
sequence of [WC] stars corresponds to an evolutionary sequence from
late to early types, ending with the PG 1159 type stars. This 
suggestion was based on the analysis of
the stellar atmospheres of [WC] stars (Hamann 1997;
 Leuenhagen \& Hamann 1998) and on the
nebular properties of the WRPNe (Acker et al. 1996).
 G\'orny \& Tylenda (2000) 
provided further observational arguments for the existence of
such an evolutionary sequence.

Tylenda et al. (1993) defined a category of ``weak emission line stars" 
(hereinafter WELS), whose emission lines have much
lower equivalent widths than the bona fide [WR] stars. Par\-tha\-sa\-ra\-thy
et al. (1998) claim that WELS are
an intermediate stage between [WC] and PG\,1159 stars.

This, however, is not the end of the story. First, not
all WRPNe belong to proposed evolutionary sequence.
  For example 4 WRPNe with late type
central stars, namely M\,4-18,
He\,2-459, He\,2-99 and NGC\,40, were considered
 to have too low densities for such a
scenario, and it
has been suggested that they may have a different evolutionary status.
\footnote{According to our data, NGC\,40, M\,4-18 and He\,2-459 could 
actually belong to the sequence. See Sect. 4.}
Second, the carbon abundance
in the wind of [WC] stars decreases with decreasing spectral
type (Hamann 1997), while the reverse is expected, at least from simple
considerations.  Third, Pe\~na et al. (1998) and de Marco \& Crowther
(1999) have
shown that PNe with nuclei of same [WC] type may have very different
nebular properties (morphologies, abundance ratios, etc.) suggesting
that stars with quite different initial masses can pass through the
same [WC] stage. Finally, the Wolf-Rayet phenomenon is known to be variable,
at least in a few cases (Pe\~na et al. 1997a; Werner et al. 1992).
Thus, it is not obvious that the stars now appearing as [WC 2-3] were
of [WC-late] type before.

In order to provide a homogeneous data set for studying
the nature and evolution of WRPNe, in 1995 we
started a program of systematic observations of WRPNe, obtaining
high resolution spectroscopic data
of the nebulae and their nuclei. One of the advantages of high
resolution is to safely deblend the nebular and stellar lines. This is
essential for compact nebulae surrounding late type stars, where the
stellar lines might otherwise affect the nebular diagnostics. In
addition, this allows to estimate expansion velocities of the nebular
envelopes.
The first results of this effort have been
presented in Pe\~na et al. (1998), where a sample of high excitation
nebulae, ionized by [WC\,2-3] central stars were analyzed by computing
detailed photoionization models for each object. In this work, we
report observational data for 29 additional PNe with nuclei of all spectral
types, including a few WELS.

In Section 2, we present our sample, describe our observations of
the nebulae and the central stars and derive the usual
plasma diagnostics (electron densities
and temperatures, abundance ratios).
In Section 3, we discuss the
systematics of the plasma properties in our sample. In Section 4,  we
examine the relation between the plasma properties and
the [WC] type of the central stars. In Section 5, we present and discuss the
expansion velocities of our sample. The main conclusions of this work
are summarized in Section 6.

\section{Observations}

We selected for observation a sample of 23 WRPNe from the list of
Tylenda et al.  (1993).  Together with the 5 WRPNe with early-type
central stars presented in Pe\~na et al (1998), this constitutes a
sample of 28 WRPNe.

We also observed 6 PNe ionized by WELS (NGC\,6629, NGC\,6578, 
NGC\,6567, NGC\,6543, IC\,5217, Cn\,2-1) from the list of
Tylenda et al.  (1993).
WELS actually represent a mixed bag, since a few of them are known to
present hydrogen features in their spectra (from our list these are 
NGC\,6629 and NGC\,6543, respectively classified as Of(H) and Of WR(H) by M\'endez 1990).
  For simplicity, in the following, we will use
the term WRPNe for all the PNe of our sample, regardless of whether
they are excited by [WR] stars or WELS, unless explicitly stated.

Our sample is not a complete one, in that
it does not reflect the true proportion of nuclei of various spectral
types, and should not be used to infer such things as stellar lifetimes
for example.
The aim was rather to obtain a homogeneous data set
for the entire range of [WC] spectral types, from [WC\,11] to [WC\,2].

The objects of our sample are presented  in Table 1, which
gives their PN G numbers from the Strasbourg-ESO catalogue (Acker et
al. 1992) as well as their usual names.

\begin{table*}
\caption{Log of observations$^{(1)}$.}
\begin{flushleft}
\begin{tabular}{lll}
\hline
& &\\
  PN G & Main Name  & Observing dates (exposure times in minutes)\\ 
  \hline \\ 
 001.5-6.7& SwSt 1 & 970804(15,10,5)\\
 002.4+5.8& NGC 6369&950729(15,15), 950731(15,15), 960615(15,15,15)  \\
 002.2-9.4& Cn 1-5  & 960617(15,15,15)    \\
 003.1+2.9& Hb 4    & 960614(15,10,10,10,15)     \\
 004.9+4.9& M 1-25  & 960617(15,15,15)    \\
 006.8+4.1& M 3-15  & 960617(15,15,15) \\                   
 009.4-5.0& NGC 6629& 970804(10,15)   \\
 010.8-1.8&NGC 6578 & 960614(15,15,15)   \\
 011.9+4.2&M 1-32  & 960614(15,15,15)      \\
 011.7-0.6&NGC 6567& 970804(10,10)  \\
 012.2+4.9&PM 1-188& 970803(15,15)    \\              
 017.9-4.8&M 3-30  & 960617(15,15)     \\
 027.6+4.2&M 2-43  & 970617(15,15,15)     \\
 029.2-5.&NGC 6751 &  970803(15,15) \\                  
 048.7+1.9&He 2-429 & 991004(15,15)    \\            
 061.4-9.5&NGC6905&950729(15,15,15,15), 950731(15,15,15), 960614(15,15,10,10)\\ 
 064.7+5.0&BD+30 3639& 970804(10,5,2), 991004(2,2,1.5)  \\
 068.3-2.7&He 2-459& 970803(15,15), 991004(15,15)        \\                  
 089.0+0.3&NGC 7026& 950730(10,15,10,1), 950731(15,15,15), 981213(15,8)  \\
 096.4+29.9&NGC 6543& 950730(2,10,10,2), 950731(10,10,10), 960615(2,2,4,4) \\
 096.3+2.3&K 3-61  & 991004(15,15)  \\
100.6-5.4&IC 5217 & 991004(10,5)   \\
120.0+9.8&NGC 40  &  981212(15,15) \\                    
130.2+1.3&IC 1747 & 950730(15), 950731(15,5), 981213(1,15)  \\
144.5+6.5&NGC 1501& 981213(15,15), 991004(15,15)  \\
146.7+7.6&M 4-18  & 981213(15,15)\\
161.2-14.8&IC 2003 & 991004(10,10,7)   \\
189.1+19.8&NGC 2371-72& 981214(15,15,15,15)    \\                                       
243.3-1.0&NGC 2452 $^2$& 981212(15,15), 981213(15,15)   \\
278.8+4.9&PB 6 $^2$   &       \\                    
278.1-5.9&NGC 2867 $^2$&       \\                    
286.3+2.8&He 2- 55 $^2$&       \\                    
352.9+11.4&K 2-16  &  960614(15,15,15), 970803(15,15) \\   
356.2-4.4&Cn 2-1  & 970804(15,15)  \\
\hline 
\end{tabular}
\end{flushleft}
$^{(1)}$Set-up July 1995: 1024x1024 23$\mu$ CCD, 3600--6700 \AA, slit 4''x13'',  res. $\sim$ 0.3 \AA\\
Set-up June 1996: 1024x1024 23$\mu$ CCD, 3500--6650 \AA, slit 4''x13'', res. $\sim$ 0.3 \AA\\
Set-up Aug.1997: 2048x2048 19$\mu$ CCD, 3360--7360 \AA, slit 4''x13'', res. $\sim$ 0.2 \AA\\
Set-up Dec.1998: 2048x2048 19$\mu$ CCD, 3360--7360 \AA, slit 4''x13'', res. $\sim$ 0.2 \AA\\
Set-up Oct.1999: 2048x2048 19$\mu$ CCD, 3360--7360 \AA, slit 4''x13'', res. $\sim$ 0.2 \AA\\
$^2$ objects observed at CTIO 1994/12/30; see Pe\~na et al. (1998). \\
\end{table*}

Several observing runs have been performed at the Observatorio
Astron\'omico Nacional, San Pedro M\'artir, B.C., M\'exico, with the
2.1-m telescope and the REOSC echelle spectrograph in the high
resolution mode. The general characteristics of the spectrograph have been
reported by Levine \& Chakrabarty  (1994). The spectral range covered
is wide enough to obtain the main diagnostic line ratios
and the spectral resolution is always better than 0.3 \AA\ per pixel.
Table 1 shows the log of observations and describes the
instrumental setup, the wavelength range and spectral
resolution for each run.

Bright spectrophotometric standard stars from the list by Hamuy et al.
(1992) were observed in all the runs for flux
calibration. A Th-Ar lamp was used for wavelength calibration in all the
spectral ranges and a tungsten bulb was used for flat-fielding.
Data reduction was performed using the IRAF\footnote{IRAF is distributed
by  NOAO, which is operated by AURA, Inc., under contract with the NSF.}
reduction package.

As Table 1 shows, 
for each PN, we have at least two consecutive observations in each
run, with exposure times allowing a good signal-to-noise in the weak lines.  
In addition, for bright objects, short exposures were obtained in order
to avoid saturation in the
most intense lines. The exposure times vary
from 2 to 15 minutes, depending on the object.
A number of PNe were observed at various epochs. The slit position was always
  centered at the central star except for some extended objects where several
  positions were observed. In this paper, we only present data for the
  central position, unless otherwise stated (see Table 2).

  After extraction and calibration of the spectra, we found that the
  differences between spectra obtained in the same night, with the same
  slit dimensions, are smaller than 5\% (under photometric conditions).
Spectra from different observing runs however show larger differences.
We estimated the errors in the line intensities from the measured
signal-to-noise and the differences between the various observing runs.

\subsection{Stellar data and classification}

The spectral types of [WC] central stars are taken from recent
literature and are listed in Table 2 (with references indicated in
the footnotes).  For WELS for which no spectral type was available,
we assigned a tentative type
using criteria based on the classification scheme of [WC] stars by van
der Hucht et al.  (1981), M\'endez \& Niemela (1982) and Hu \& Bibo
(1990).  This scheme is based on the relative strength of the optical
\ion{C}{iv} 5805 and \ion{C}{iii} 5695 lines and the \ion{O}{v} 5598,
\ion{O}{vi} 5290 and \ion{O}{vii} 5670 lines.  Crowther et al.  (1998)
proposed a unified and more quantitative classification scheme of WC
and WO stars, which is independent of the C/O ratios in the
star atmospheres but requires good signal-to-noise in the diagnostic
lines. We considered the option of using this scheme for all our
objects, but because of the high resolution of our spectra, which
spreads the stellar features over many pixels, we were able to classify
accurately only
the stars brighter than about $m_{V}$=16.  In any case, for those objects
for which several classifications are available, the proposed spectral types
differ by one unit at most.
It must be recalled that
any spectral classification of [WC] stars orders the objects with
respect to one parameter only, basically the ionization structure of
the atmosphere.  This ionization structure depends primarily on the
temperature of the star, but other parameters such as the wind density
do play a role.  So, even in the Crowther et al. (1998) classification
scheme, two stars of different temperatures may have the same spectral
type if they differ in the other wind properties.

We also list in Table 2 the stellar visual magnitudes $m_{V}$ derived
from the stellar continuum fluxes at 5480 \AA\ measured in
our spectra (after subtracting the nebular continuum using the
expressions given by Pottasch 1984). For stars brighter than
$m_{V}$=15, the uncertainties in our measurements
are about $\pm$0.3 mag considering that a small fraction of the light
could have escaped from the narrow slit.  Fainter magnitudes have much larger
 uncertainties. For the bright stars, our values are in
general in agreement with the stellar magnitudes reported in Acker et al.
(1992). There are however some discrepancies larger than 0.5 mag. For instance, 
in our data Cn\,1-5 is fainter, NGC\,6567 is brighter, NGC\,6551 is brighter, 
K\,2-16 is fain\-ter, etc.,  which are possibly indicating that stars could 
be variable. Our present data do not allow us to analyze the origin of the 
discrepancies or the possible stellar variations. Better photometric data  
are required  for such an analysis.

\subsection{Nebular data}

  The observed
calibrated fluxes were corrected for reddening according to the
expression:

\begin{equation}
{\rm log} ~I(\lambda) ~/ ~I({\rm H}\beta) = {\rm log} ~F(\lambda) 
~/~F({\rm H}\beta) + c({\rm H}\beta) \times f_\lambda
\end{equation}

\noindent where $I(\lambda$) and $F(\lambda$) are the dereddened and
observed fluxes respectively, $c(\rm H\beta$) is the logarithmic
reddening correction at H$\beta$, and $f_\lambda$ is the reddening
law.  We employed the reddening law by Seaton (1979).  The value of
$c(\rm H\beta$) was derived for each object from the Balmer decrement,
by considering case B recombination theory (Hummer \& Storey 1987).

Dereddened intensities for the most important lines, relative to
H$\beta$, are listed in Table 2. The 5 objects
already presented in Pe\~na et al. (1998) are also listed in Table 2,
where additional data not reported before were included.

The intensities of the strong lines 
(observed flux larger than 
 $\sim  10^{-14}$ erg cm$^{-2}$ s$^{-1}$) are generally accurate within
10\%. Those with observed fluxes in the range  
$ 1 - 5\times 10^{-15}$ erg cm$^{-2}$ s$^{-1}$  
are accurate within
20 -- 30\%. The data marked with a colon have uncertainties larger than 50\%.
The line ratios that are used to derive
electron temperatures and densities are explicitly listed at the bottom 
of Table 2, together with their uncertainties.

Table 2 also gives the values
of $c(\rm H\beta$) derived from our observations and
  the observed fluxes in H$\beta$.
These H$\beta$ fluxes  are not of photometric quality but are given as an
indication of the brightness of the objects. We also list in Table 2
the nebular diameters, taken from the Strasbourg-ESO PN catalogue (Acker et al. 1992).

\subsection{Plasma diagnostics}

The plasma diagnostics were performed from the emission line ratios in
a standard way, using the same atomic data as listed in Stasi\'nska \&
Leitherer (1996).

The wide wavelength range of our observations allowed us to
determine the electron densities and electron temperatures from several
line ratios.

  Electron densities were derived from [\ion{O}{ii}] 3726/3729, [\ion{S}{ii}]
6717/6731
and [\ion{Ar}{iv}] 4711/4740, electron temperatures from [\ion{O}{iii}]
4363/5007 and [\ion{N}{ii}] 5755/6583. The densities used in the
derivation of the electron temperatures were those deduced from
[\ion{O}{ii}] 3726/3729, when available, [\ion{S}{ii}]
6717/6731 otherwise. At the densities encountered in our sample, only
  the [\ion{N}{ii}] 5755/6583 temperature is dependent on the densities,
  and only to a small amount.

The derived electron temperatures and densities are listed in Table 3,
together with the error bars based on the uncertainties in the line
ratios given in Table 2.

Ionic abundances were then obtained, using $T_{\rm e}[\ion{O}{iii}]$ for
the high ionization species and  $T_{\rm e}[\ion{N}{ii}]$ (when available) for the low
ionization ones. All the ionic abundances were computed using the electron
densities derived from [\ion{O}{ii}] 3726/3729
($n_{\rm e}$ from [\ion{S}{ii}] 6717/6731 was used when the former was not available). 
The value of the density is not important for most of the ions
we are interested in, except O$^{+}$.

Elemental abundance ratios were computed from the ionic abundance
ratios using the ionization correction factors of Kingsburgh \& Barlow
(1994) (note that these ionization correction factors are
in reasonable agreement with the modeling performed for the highly 
ionized objects studied in Pe\~na et al. 1998).
The abundance ratios He/H, O/H, N/O and Ne/O are given at the bottom
  of Table 3. The error bars that are listed for O, N and Ne only take into account
  the uncertainties propagated from the uncertainties in the physical
  conditions (electron temperature and density). 
  Uncertainties in the ionization
  correction factors are not included.
 It must be kept in mind that, in the case of low
  excitation objects, evidently for Sw St 1, BD+30 3639, He\,2-459 and NGC\,40, 
the helium abundances derived in
  such a way are merely lower limits, since neutral helium is not taken
  into account. It is only with a detailed photoionization modeling
   that one could put stronger constraints on the helium
  abundance in these objects, an attempt which is outside the scope of 
this paper.

\section{Systematics of nebular properties in our WRPNe sample}

\subsection{Density and temperature structure}

In this section, we compare the different density and temperature
diagnostics, with the aim of looking for systematic trends.
PNe are ordered as a function of O$^{++}$/O$^{+}$, a 
parameter which provides
an easy description of the ``excitation" of the nebula and is available
for most of the objects of our sample. The few exceptions are
  PM\,1-188,
 He\,2-459, K\,3-61, NGC\,1501 and M\,4-18.

Figure  1-a shows the ratio $n_{e [\ion{S}{ii}]}$ / $n_{e[\ion{O}{ii}]}$ as a 
function of
O$^{++}$/O$^{+}$. PNe with [WC] central stars
are represented by filled circles, PNe around WELS are represented by
open circles (as will be the case for all our observational
diagrams).  Although the two determinations
of $n_{e [\ion{S}{ii}]}$ and $n_{e[\ion{O}{ii}]}$ are fairly similar, 
the diagram shows a tendency for
$n_{e [\ion{S}{ii}]}/n_{e[\ion{O}{ii}]}$ to be larger than one for WRPNe of low
excitation, and marginally equal to one for WRPNe of high excitation.
This tendency is the same if we use the [\ion{S}{ii}] collision strengths from
Keenan et al. (1996) instead of those from Cai \&
  Pradhan (1993) that were used here.
  
In principle, one expects [\ion{S}{ii}] lines to be emitted in regions of lower
ionization than [\ion{O}{ii}] lines, and this is confirmed by photoionization
models.  One explanation of this trend could be that, for low
excitation nebulae at least, which are probably ionization bounded,
the density is increasing outwards, towards the ionization front. For
density bounded nebulae (as is probably the case for
many high excitation objects), the emission in the [\ion{S}{ii}] and [\ion{O}{ii}]
lines is largely due to trace ions present in the O$^{++}$ zone, and the
[\ion{S}{ii}] and [\ion{O}{ii}] densities are expected to be equal.

Figure 1-b shows $n_{e [\ion{Ar}{iv}]}$ / $n_{e[\ion{O}{ii}]}$ as a function of
O$^{++}$/O$^{+}$. In the nebulae where $n_{e [\ion{Ar}{iv}]}$ can
be derived, i.e. in the high excitation ones only, this ratio is
generally compatible with 1 and, statistically, indicates no tendency for
a density gradient.

It would be interesting to have a control sample of non-WRPNe observed
with the same equipment to see
whether ordinary PNe show a similar behavior.

Figure 1-c shows the ratio  $T_{\rm e}[\ion{O}{iii}]$ / $T_{\rm e}[\ion{N}{ii}]$ of the 
electron temperatures
derived from the [\ion{O}{iii}]4363/5007 and [\ion{N}{ii}]5755/6583 ratios
respectively,  as a function of
O$^{++}$/O$^{+}$. We
note a tendency for $T_{\rm e}[\ion{O}{iii}]$ / $T_{\rm e}[\ion{N}{ii}]$ to decrease as the
excitation increases, with values significantly lower than 1 at the
high excitation end.

Simple photoionization models at metallicities around half solar,
typical of the objects
of our sample (see next section), behave quite differently.
$T_{\rm e}[\ion{O}{iii}]$ / $T_{\rm e}[\ion{N}{ii}]$ increases steadily with stellar effective
temperature as soon as T$_{*}$ becomes larger than 50,000K,
and exceeds one for T$_{*}$ $>$
100,000K. The reason why the electron temperature increases outwards at
low T$_{*}$ is that the
most energetic stellar photons are absorbed in the outer regions, and
that the ions found there are less effective for the cooling.
At high T$_{*}$, the energy gains are larger, inducing a
  higher electron temperature. Then, the ions in the outer zone become more
effective for the cooling than the ions in the O$^{++}$ zone, resulting in
a negative outward temperature gradient in the nebula. In any case,
the temperature gradient shown by the models is very mild. The
  $T_{\rm e}[\ion{O}{iii}]$ / $T_{\rm e}[\ion{N}{ii}]$ versus
O$^{++}$/O$^{+}$ plot will be investigated further, both
theoretically and observationally. We plan to
obtain observations of a control sample of non-WRPNe with a
similar equipment, to see if non-WR nebulae behave in a similar
way (the publicly available data on large samples of PNe do not allow us
to analyze this question with confidence, because $T_{\rm e}[\ion{O}{iii}]$
and $T_{\rm e}[\ion{N}{ii}]$ are rarely available together and with good accuracy
for the same object).
For the moment, one can speculate that the low
$T_{\rm e}[\ion{O}{iii}]$ / $T_{\rm e}[\ion{N}{ii}]$ we see in our WRPNe sample could be due 
to a strongly
inhomogeneous structure (but this does not seem supported by our
measurements of the $n_{e [\ion{Ar}{iv}]}$ / $n_{e[\ion{O}{ii}]}$ ratio), or to 
additional heating by shocks or
turbulence for example.

\begin{figure*}
\resizebox{17cm}{!}{\includegraphics{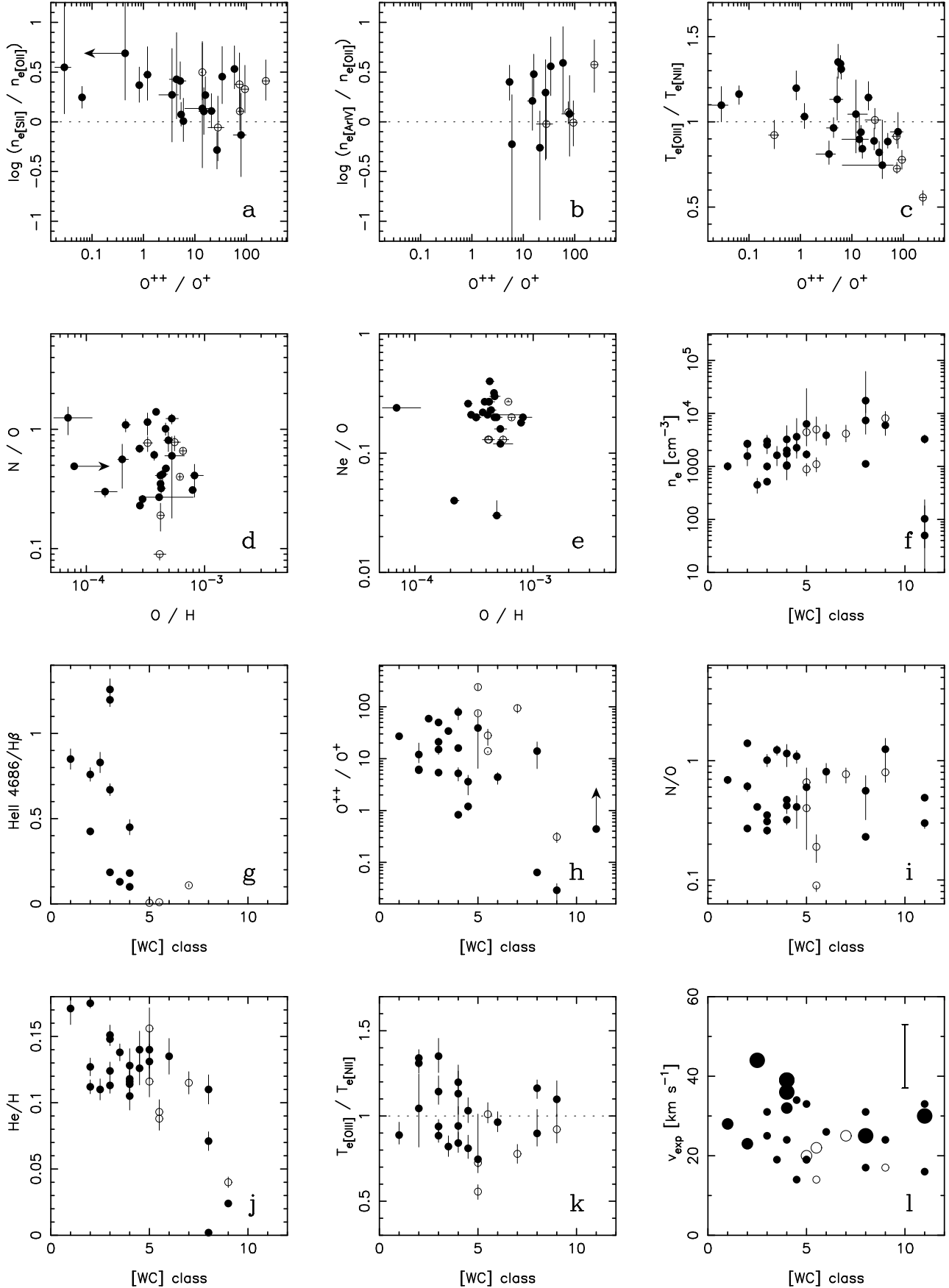}}
\caption[]{The behavior of different nebular characteristics. Filled  
circles: PNe with [WC] nuclei, open circles: PNe with WEL stars. The [WC] types are defined with an uncertainty of one unit.
The error bars in the computed quantities are indicated. In Panel l, we show the
typical error bar corresponding to the measurement of the FWHM of the H$\beta$ line;
large symbols indicate PNe with split line profiles; intermediate symbols, 
PNe with complex profiles and small symbols, PNe with Gaussian profiles.}\label{Fig1}
\end{figure*}

\subsection{Abundance ratios}

Regarding the O/H ratios in our sample of WRPNe, we find that they spread
over almost a factor 10
with a mean value $\sim 4\times 10^{-4}$. Such a distribution is 
comparable to the
distribution found for example in the sample of PNe studied by
Kingsburgh \& Barlow (1994), which contains only about 10\% of WRPNe.
  One reason for such a large
range is simply that the objects are located at various galactocentric
distances, and that PNe are known to show an oxygen abundance gradient
in the galaxy (Maciel \& K\"{o}ppen 1994). That there is no
significant difference in the O/H distributions between our sample and
that of Kingsburgh \& Barlow (1994) probably reflects the fact that
the spatial distributions of WRPNe and ordinary PNe in the Galaxy are
similar, as indicated by the similar distribution of coordinates and
proper motions (Acker et al. 1996).

Diagrams relating abundance ratios in planetary nebulae are commonly
used to cast some light on the nucleosynthesis and dredge-up
processes in the 
progenitor stars (e.g. Henry 1990; Perinotto 1991;
Groenewegen et al. 1995).

Figure 1-d  shows N/O as a function of O/H for our sample of WRPNe. A
large spread in N/O is seen, with values in N/O ranging over more than
a factor 10. Such a spread is commonly observed in
general samples of PNe (e.g. Torres-Peimbert \& Peimbert 1977;
Kingsburgh \& Barlow 1994; Leisy \& Dennefeld 1996). It should be
noted that N/O is derived from the [\ion{N}{ii}]6584/[\ion{O}{ii}]3727 intensity
ratio, and is strongly affected by errors in the flux
calibration,  in the dereddening procedure
 and in the adopted electron densities.
Nevertheless, the spread is much larger than the uncertainties. Some
objects clearly show enhanced nitrogen abundances, which indicates
that secondary nitrogen has been dredged-up to the surface in the
central star progenitor. Others show quite modest N/O ratios. One
interpretation is that the progenitors of [WC] stars had quite
different initial masses, similarly to the progenitors
of the central stars of ordinary PN, since
the amount of dredged-up nitrogen in a post-AGB star depends on the
progenitor's mass. However, reality could be more complex,
as nitrogen yields depend also on the intensity of mass-loss on the AGB
(Forestini \&
Charbonnel 1997; van den Hoek \& Groenewegen 1997),
and progenitors of [WC]
central stars might experience different mass-losses than progenitors
of ordinary central stars.

Figure 1-e shows Ne/O as a function of O/H in our sample of WRPNe.
Compared to the N/O values, the Ne/O ones are much less scattered.
  This is also what is found
in other samples of PNe (Henry 1990; Kingsburgh \& Barlow 1994;
Stasi\'nska et al. 1998) and is taken as evidence that oxygen and neon
abundances are not modified by nucleosynthesis during the life of the
progenitors. However, we note that two objects, whose neon abundances
are measured for the first time (M\,1-25 and M\,1-32) have Ne/O
smaller by about a factor 10 than the rest of our WRPNe (both these objects
have  O$^{++}$/O$^{+}$ $>$ 1, therefore one does not expect the ionization
correction factor to introduce a significant error in the
determination of Ne/O). This is reminiscent of the planetary nebula H 4-1
in the Galactic halo, which has an Ne/O ratio of 1.5~10$^{-2}$ 
(Howard et al. 1997),
and for which there seems to be no explanation so far. Note that, apart from 
their unusual Ne/O ratio, M\,1-25 and M\,1-32 do not stand out particularly
 in any of the diagram in Fig. 1, except that M\,1-32 lies at the lower limit
 of expansion velocities (see Sect. 5 and Appendix).

\section{Is the proposed evolutionary [WC] sequence reflected in the nebular
  parameters?}

We now analyze the nebular parameters as a function of the spectral type
of the nucleus (panels f through l in Fig. 1)\footnote {In panels f through l, 
M\,3-30, whose central star type is OVI, is represented with an abscissa of 1.
Cn\,2-1, whose central star type is undetermined, is not represented.}. 

Fig. 1-f shows the electron density as a function of the [WC] spectral
type. In this figure, the electron density is derived from the [\ion{O}{ii}] ratio
(except in 7 cases where this ratio was not available and the density
was derived from the [\ion{S}{ii}]). One expects $n_{e[\ion{O}{ii}]}$ to be
  a better indicator of the overall nebular density than $n_{e[\ion{S}{ii}]}$,
  since S$^{+}$ occupies a smaller volume than O$^{+}$.
$n_{e [\ion{Ar}{iv}]}$ would probably be even
better, but this data is available only for a few objects.
We find that the density decreases from late to early
[WC] types. The slope is not as large as seen in the figure of
  G\'orny \& Tylenda (2000), who used the $n_{e[\ion{S}{ii}]}$ densities,
  but the trend is clearly there.
  In our sample,
  we do not have many of the objects that, in the G\'orny \& Tylenda (2000) compilation,
  have  $n_{e[\ion{S}{ii}]}$ $>$
  $10^{4}$ cm$^{-3}$. Also, since we
  use  $n_{e[\ion{O}{ii}]}$  rather than  $n_{e[\ion{S}{ii}]}$,
  BD+30 3639 is plotted with a density smaller than  $10^{4}$ cm$^{-3}$.

  Acker et al. (1996) and G\'orny \& Tylenda (2000) interpreted the
  decrease in electron density with decreasing [WC] spectral type
   as evidence for an evolutionary sequence
  from late to early type [WC].
  Of the four objects that G\'orny \& Tylenda (2000) found to lie
  below the sequence (see the introduction),
  we have observations of NGC\,40, M\,4-18 and He\,2-459. In our
  measurements, we find for He\,2-459 a much larger density than used
  by these authors (17,000 instead of 3,600 cm$^{-3}$), therefore this object is
  well within the sequence (see also Appendix).  As for NGC\,40 and M\,4-18, they lie only marginally below the sequence. 

On the other hand, two
  objects of [WC\,11] type, K\,2-16 and PM\,1-188, which
  had no previous determination of the
  density, are clearly well below the sequence. These are then candidates
  for being PNe with central stars having experienced a late helium flash and
  returned to the AGB for a ``born again" evolution. Alternatively, the
  stars may have been evolving very slowly off the AGB, either
  because of their low mass, or because of a very long transition time
  between the tip of the AGB and the beginning of the fast wind.
Unfortunately, these objects are of very low excitation, so that there is only 
little relevant information in our spectra that could give some additional 
clues to their evolutionary status. PM\,1-188 appears only in panels f and l 
of Fig. 1.
 K\,2-16 appears also in panels a, d, h and i (as a limit in the first
 three of them), and does not 
stand out particularly with respect to the other nebulae of our sample.
  In principle, it should be possible to distinguish
  between a ``born again" scenario and one of slow stellar evolution. In the
  first case, the ionization stage of the nebula should be higher than the
  equilibrium value corresponding to the effective temperature of the
  central star, at least if the helium flash occurred recently (about a
  few hundred years ago). The spectra of K\,2-16 and PM\,1-188 do not
  seem to indicate that the nebular gas is presently recombining from a
  more ionized stage, since no \ion{He}{ii} lines are observed,
  favoring the hypothesis of a very slow evolution
  of the central star. However, a more careful theoretical and
  observational study of these objects
  would be warranted.

  Note that, in Fig. 1-f,  PNe that are ionized by
  WELS have intermediate densities as compared to the  WRPN
  sample. If, as suggested by Par\-tha\-sa\-ra\-thy et al. (1998), WELS were
  the descendents of [WC] early stars, they should be surrounded by more
  diluted nebulae.  The fact that WELS are inside the [WC] sequence is 
intriguing. One explanation could be that [WC] and WELS are transient stages of the
  same stars.

Fig. 1-g shows that the nebular \ion{He}{ii} 4686 line is seen only for early type [WC]
stars, with the \ion{He}{ii} 4686/H$\beta$ ratio tending to increase with decreasing
spectral type. Qualitatively, this is expected, since spectral types
are related with stellar effective temperatures. Differences in
slit coverage, nebular masses, and stellar wind densities
probably explain the scatter.

Fig. 1-h shows O$^{++}$/O$^{+}$ as a function of [WC] class.
Only a weak trend is present.
Actually, one does not expect a perfect correlation even between
O$^{++}$/O$^{+}$ and stellar effective temperature, since the stellar
luminosity, the nebular
mass, its average density and its density structure all affect the
value of O$^{++}$/O$^{+}$. Therefore, the large dispersion seen  in
O$^{++}$/O$^{+}$  at a
given [WC] spectral type is not surprising.

Fig. 1-i shows N/O as a function of [WC] class. One can note the large
scatter in N/O at a given [WC] spectral type, reinforcing the
suggestion that WRPNe have a range in progenitor
masses.

  Fig. 1-j displays He/H as a function of [WC] class. As noted
  previously, the points with the lowest values of He/H correspond to
  lower limits, since no account has been made for the presence of
  neutral helium to derive the total helium abundance. Figure 1-j shows
  that these points represent objects with the latest, and therefore
  coolest [WC] central stars,
  which are indeed expected to
  produce a He$^{+}$ zone less extended than the H$^{+}$ zone. For
  intermediate and early type [WC] spectral classes, the helium
  abundances are reliable and we find a larger scatter than expected
  from observational uncertainties (which are of about 10 - 20\% for each
  object). The
  contribution of the helium-rich wind from the [WC] star to the mean
  He/H ratio observed in a nebula is expected to be rather small if we
  adopt a typical mass-loss rate of 10$^{-6}$~M$_{\odot}$ yr$^{-1}$ and a lifetime of
  10,000 yr. Therefore, the observed scatter in He/H reflects differences in
  the chemical composition of the star atmospheres while on the AGB,
  likely due to differences in progenitors masses.

Fig. 1-k shows $T_{\rm e}[\ion{O}{iii}]$ / $T_{\rm e}[\ion{N}{ii}]$ as a function of [WC] 
class. There is
no clear trend in this figure, as compared with the trend seen in
Fig. 1c. This seems to indicate that the behavior of $T_{\rm e}[\ion{O}{iii}]$ / 
$T_{\rm e}[\ion{N}{ii}]$
  has more to do
with the nebular properties than with stellar properties or with an
evolution scenario.

Interestingly, the WELS do not stand out with respect to the WRPNe in none
of the diagrams discussed so far.

\begin{figure}
\resizebox{7cm}{!}{\includegraphics{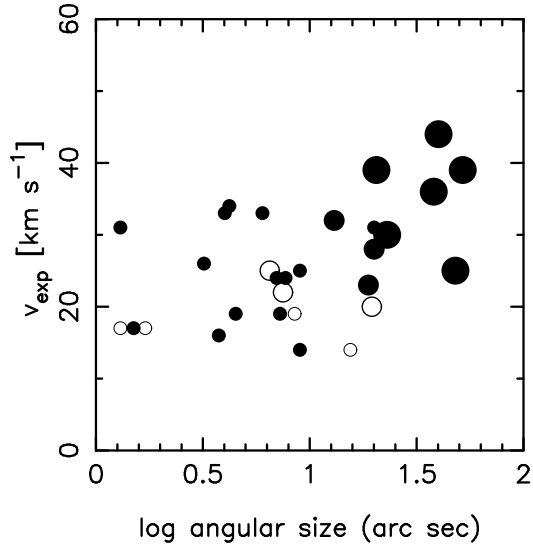}}
\caption[]{The dependance of $v_{exp}$ on the nebular diameter is shown. Symbols are the same as in Fig. 1-l.}\label{Fig2}
\end{figure}

  \section{The expansion velocities}

Velocity fields of planetary nebulae are an indicator of
  the genesis and the evolution of these
objects (Mellema 2000). Basing on data compiled from the literature,
it has been shown by G\'orny \& Stasi\'nska (1995) that, on
average, WRPNe have larger expansion velocities than non-WR PNe. This
was interpreted as possibly due to the effect of a stronger stellar wind
in the case of WRPNe.

The high spectral resolution of our observations allowed us to study
the emission line profiles in most objects of our sample.
  A detailed
description of the velocity fields, based on the study of the profiles
for lines of various ions, will be given elsewhere (Medina et al.,
  in preparation). Here we use a simple estimate of the expansion
  velocity, $v_{\rm exp}$, derived from the H$\beta$ profile, to see
  whether there is a correlation between expansion velocity and spectral
  type of the central star. Indeed, using expansion velocities compiled
  from the literature,  Acker et
  al. (1996) found  $v_{\rm exp}$ to increase with decreasing [WC] type.

  Deriving expansion velocities from line profiles is a notoriously
  difficult issue. The results depend on the adopted method
  (as shown by Gesicki \& Zijlstra 2000 using simple expanding
  photoionization models) and compiling expansion velocities from various
  sources may be misleading. Even observing all the
  objects with the same instrument and adopting the same procedure to
  derive the average expansion velocity, the result depends on the
  slit coverage and on the geometry of the source.
In our objects we have found very different nebular line profiles. Several of
them show split lines indicating the existence of
a well defined shell. For such objects, the expansion velocity was
assumed to be given by half the peak-to-peak distance in the H$\beta$ profile.
In other cases, only one resolved line component (roughly Gaussian)
was found. In some objects, spatially resolved knots or condensations were
seen. In all the cases of one component, the lines profiles were spectrally
resolved,
showing a FWHM larger than the instrumental resolution. The expansion
velocity was then assumed to be given by the half width at half
maximum (subtracting the instrumental width in
quadrature). The expansion velocities derived in such a
way for the objects of our sample are reported in Table 2, together
with an indication of the shape of the line profile (split,
Gaussian, complex).

Figure 1-l displays $v_{\rm exp}$ as function of the [WC]-type.
Objects with split lines are represented by the largest circles,
objects with Gaussian profiles with the smallest
circles. Complex profiles are indicated with
intermediate circles. As in the preceding figures, filled circles are
for true [WC], open circles are for WELS.

 From this figure, the trend of $v_{\rm exp}$ increasing with
decreasing [WC] type relies on four objects (out of 29) which have the highest
expansion velocities (NGC\,6905, NGC\,6751, NGC\,1501 and NGC\,6369).
These four objects show split line profiles. Their larger measured
expansion velocities may simply reflect an observational bias. This is
confirmed by Fig. 2, which plots  $v_{\rm exp}$ as a function of the
apparent nebular diameter. It is clear, from this figure, that the
type of line profile seen in our spectra depends on the fraction of
the nebular image that is encompassed by the slit, and that the
largest expansion velocities are found for the largest nebulae.
 From this, we conclude that there is no
clear evidence of an increase in $v_{\rm exp}$ with decreasing
spectral type, contrary to what is reported by Acker et al. (1996).
This weakens somehow the statement that most [WC] stars evolve from late to early 
types. Indeed, the naive expectation is that, in such a case,
the strong winds from the star would produce an increase of the 
nebular expansion velocity with time. This, however, 
should be checked by detailed hydrodynamical models following the 
coupled stellar and nebular evolution on a sufficiently long timescale.

Interestingly, WELS show lower values of $v_{\rm exp}$ than bona fide WRPNe of
similar angular size. Since the expansion velocity of a nebula depends
on its entire former history, this fact tends to contradict the
suggestion made in the previous section that WELS and [WC] stars might
be transient stages of the same objects. The nature of WELS is by no
means elucidated. As mentioned above, WELS probably represent a
mixed bag. One obviously would need to observe a larger
  sample of these objects and to study simultaneously the stellar and
  nebular properties.

\section{Summary and prospects}

We have obtained high signal-to-noise  high resolution spectra of a large
sample of PNe with [WC] spectral type nuclei and a smaller sample of
nebulae with weak emission line central stars. The nebulae are ionized by
stars spanning the whole range of [WC] types, from  [WC\,2] to
[WC\,11]. These ample data, treated in a consistent way, enabled
us to perform a significant statistical study of these objects.
The main results are the following.

Concerning the plasma properties of the nebulae, we found a
significant trend of the electron temperature ratio
$T_{\rm e}[\ion{O}{iii}]$ / $T_{\rm e}[\ion{N}{ii}]$ decreasing with
increasing excitation of the object as measured by the O$^{++}$/O$^{+}$
ratio. At the high excitation end, $T_{\rm e}[\ion{N}{ii}]$ is larger than
$T_{\rm e}[\ion{O}{iii}]$ by several thousands degrees. This is not expected from
simple photoionization models of planetary nebulae, and might indicate
the presence of additional heating mechanisms in the outer zones
(shocks or turbulence for example). Future observations of a control
sample of non-WRPNe will enable us to show whether such a behavior is
also seen in non-WR nebulae, giving some clues as to whether this
additional heating mechanism can be due to the effect of the intense
stellar winds from the central stars in the case of WRPNe.

The observed large dispersion in the N/O ratios in WRPNe indicates that
there is no preferential stellar mass for the Wolf-Rayet phenomenon
to occur in the nucleus of a planetary nebula, confirming the finding
by G\'orny \& Stasi\'nska (1995).

Two objects, though, M\,1-25 and M\,1-32, whose neon abundances
have been measured for the first time, have a Ne/O ratio of the order of 0.03,
i.e. about ten times smaller than found in other PNe. There is only
one PN known with a similarly low neon abundance: H 4-1, in the
Galactic halo (Howard et al. 1997).  There is no explanation so far
for such anomalous Ne/O ratios in planetary nebulae.

Our data set enabled us to reexamine the relation between
the nebular properties of the WRPNe and the spectral types of the central
stars. It has been suggested that the [WC] sequence corresponds to an
evolutionary sequence, from late to early type [WC]. Our plot of the
electron density versus [WC] spectral type, obtained with a
homogeneous data set, indicates that nebulae around [WC] early stars are
more evolved than nebulae around [WC] late stars, seeming to confirm 
such a view.
However,
two objects with extremely late [WC] central stars (K\,2-16 and PM
1-188) for which the
electron densities have been determined for the first time, were shown
to be of low density (well below 1,000 cm$^{-3}$). They may be similar in
nature to He\,2-99,
which was not in our sample, but was suggested by Acker et al. (1996)
to be powered by a star having experienced a late-helium flash.
Alternatively, the central stars of these low density nebulae may
have evolved on much larger
  time scales than the rest of PN nuclei, either due to their unusually
  small core mass or to an exceptionally long transition time from the
  tip of the AGB. Further observational and theoretical studies of
  these objects could enable one to distinguish between the various
  interpretations.

  Our high resolution data allowed us to estimate the expansion
  velocities for almost all the PNe from our sample. Contrary to the
  claim by Acker et al. (1996), we do not find evidence for an increase
  in $v_{\rm exp}$ with decreasing spectral type. This, then, casts some
  doubt on the hypothesis that the [WC] stars evolve from late to
  early spectral type, since one would expect the expansion velocity to increase
  with time. However, a proper determination of the velocity fields in
  our planetary nebulae would require a more thorough analysis
than performed here. Also, grids of detailed hydrodynamical models similar to the 
ones by Frank (1994)  or Mellema (1994) but
including cases of massive stellar winds would be necessary to test under what
conditions exactly one expects
  the expansion velocities to increase with time.

  Our sample includes 6 objects with weak emission
  line central stars (WELS).  It has been suggested
    by Par\-tha\-sa\-ra\-thy et al. (1998) that such stars are the
evolved counterparts of early-[WC] stars (this, of course, cannot hold
  for the 2
WELS of our sample that have been reported to present clear hydrogen
features).
   Most of our WELS are of intermediate [WC] types
   (when classified following criteria on
  intensity or equivalent width ratios of [WC] emission lines).
  In general, we find that the nebulae ionized by WELS follow
  the same tendencies as bona fide WRPNe.
  However, they show systematically lower expansion
velocities than WRPNe of similar angular size (taking into account
the angular size is important for the comparison of the expansion
velocities, as we have shown). If one assumes that expansion velocities
should be larger in the case of stronger central star winds,
  then the natural conclusion is that the WELS of our sample
were not bona fide [WC] stars in the past. Such a conclusion is supported
by the fact that the densities in nebulae ionized by WELS are not
particularly low.
Obviously, a much larger sample of
WELS should be investigated before one can draw meaningful
conclusions about the status of WELS.

It is interesting to compare the statistical properties of WRPNe in our
Galaxy with those in the Magellanic Clouds (MC). There are four well 
established [WC]
in the LMC (N\,110, N\,133, N\,203
and MA\,17) and two in the SMC (N\,6 and Ln\,302).
This sample was studied in detail by Pe\~na et al. (1997b).

The object SMC-Ln\,302 is excited by a [WC\,8] nucleus and, similarly to the
galactic [WC\,8] NGC\,40, presents an electron density of about 1000~cm$^{-3}$.
That is, an apparently
evolved nebula around a rather cool [WC].
All the other Magellanic Clouds [WC] stars have spectral types ranging from
[WC\,4] to [WC\,7], and are surrounded by dense nebulae ($n_{e[\ion{S}{ii}]}$
$>$ $10^{4}$ cm$^{-3}$).
In the Galactic sample, only half of the [WC] stars are
in the [WC\,4] to [WC\,7] range (Tylenda et al. 1993; G\'orny \& Tylenda
2000) and  their [\ion{S}{ii}] densities are mostly lower than 10$^{4}$cm$^{-3}$.

Besides the striking difference concerning [WC] spectral type
distributions and relations with nebular densities, another noticeable
  difference regards N/O. For the objects in the Magellanic Clouds,
  the N/O ratios are found between 0.1
and 0.2 (excepting SMC-N6 which presents
  N/O about 0.65), while for the objects in our Galactic sample
  they occupy a wide range of values and are all above 0.2 (30\% objects have 
N/O $>$ 0.5).
  This offset in the distribution of N/O ratios between the Magellanic
  Clouds and the Milky Way WRPNe is similar to the one found for non
  WRPNe (see Fig. 1 by Stasi\'nska et al.  1998), where only the
luminous PNe must be considered, since the WRPNe identified in the MC
are all luminous). The fact that the distribution of N/O in MC
WRPNe is extremely narrow could be an effect of small
number statistics. In any case, the low  N/O observed in the known WRPNe of
the MCs seems to indicate that the [WC] progenitors did not undergo the
second dredge-up event (although they obviously underwent
a dredge-up phase that extracted large amounts of freshly-made carbon
up to the stellar surface) implying that they possibly were not 
massive. This is at variance with the fact that the progenitors of 
Galactic WRPNe seem to have a large range of masses.
  Notoriously, the only known WRPN in the Magellanic Clouds
  with a probably high mass progenitor
  is LMC-N66, whose central star exhibits a highly variable
mass loss,
developing (apparently episodically) intense [WN]-early features
(Pe\~na et al. 1994;  Pe\~na et al. 1997a).

  It seems clear that the differences in metallicity between the Magellanic
  Clouds and the Milky Way are largely affecting the [WR] phenomenon in central
  stars of planetary nebulae.

\begin{figure*}
\resizebox{19.5cm}{!}{\includegraphics{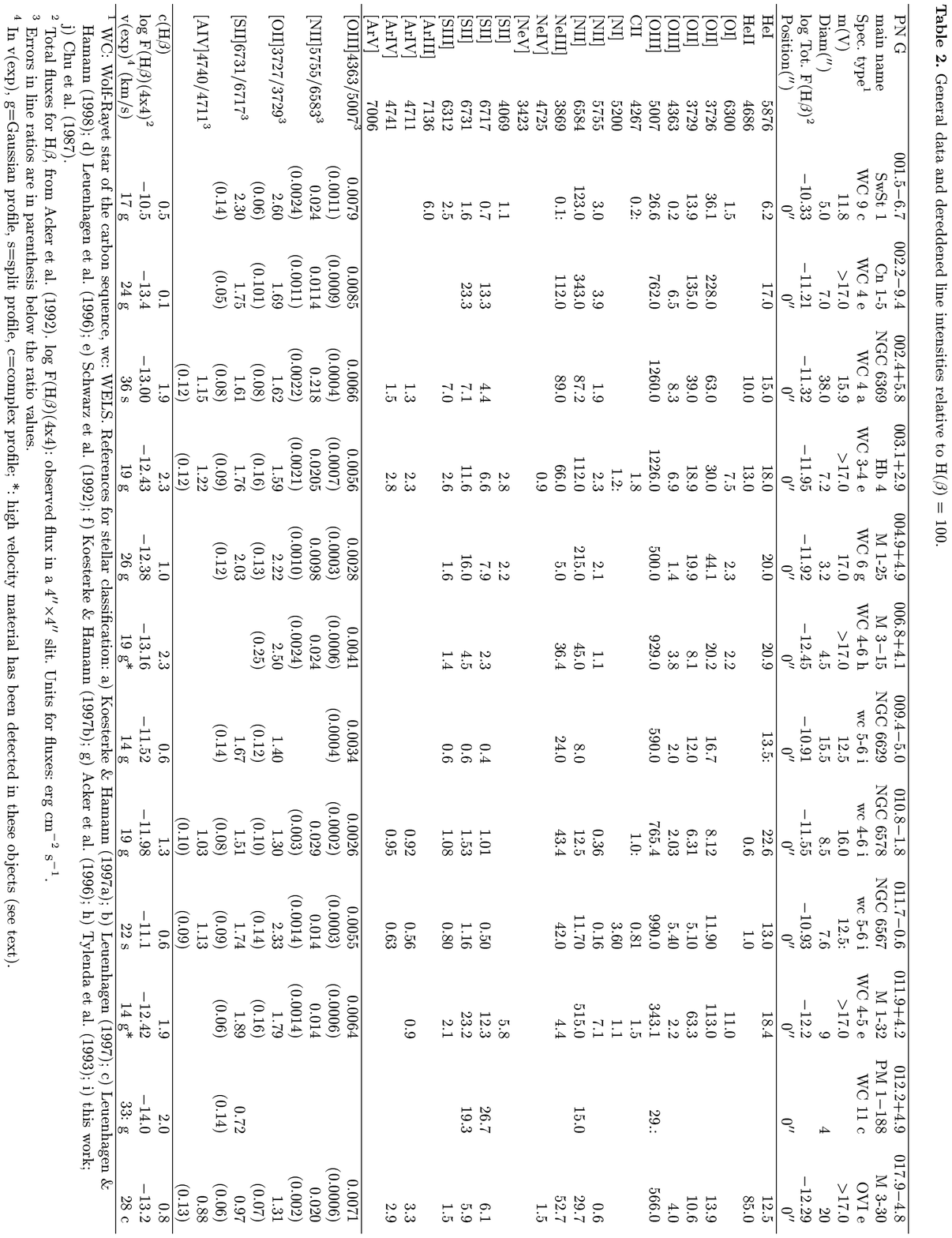}}
\end{figure*}

\begin{figure*}
\resizebox{19.5cm}{!}{\includegraphics{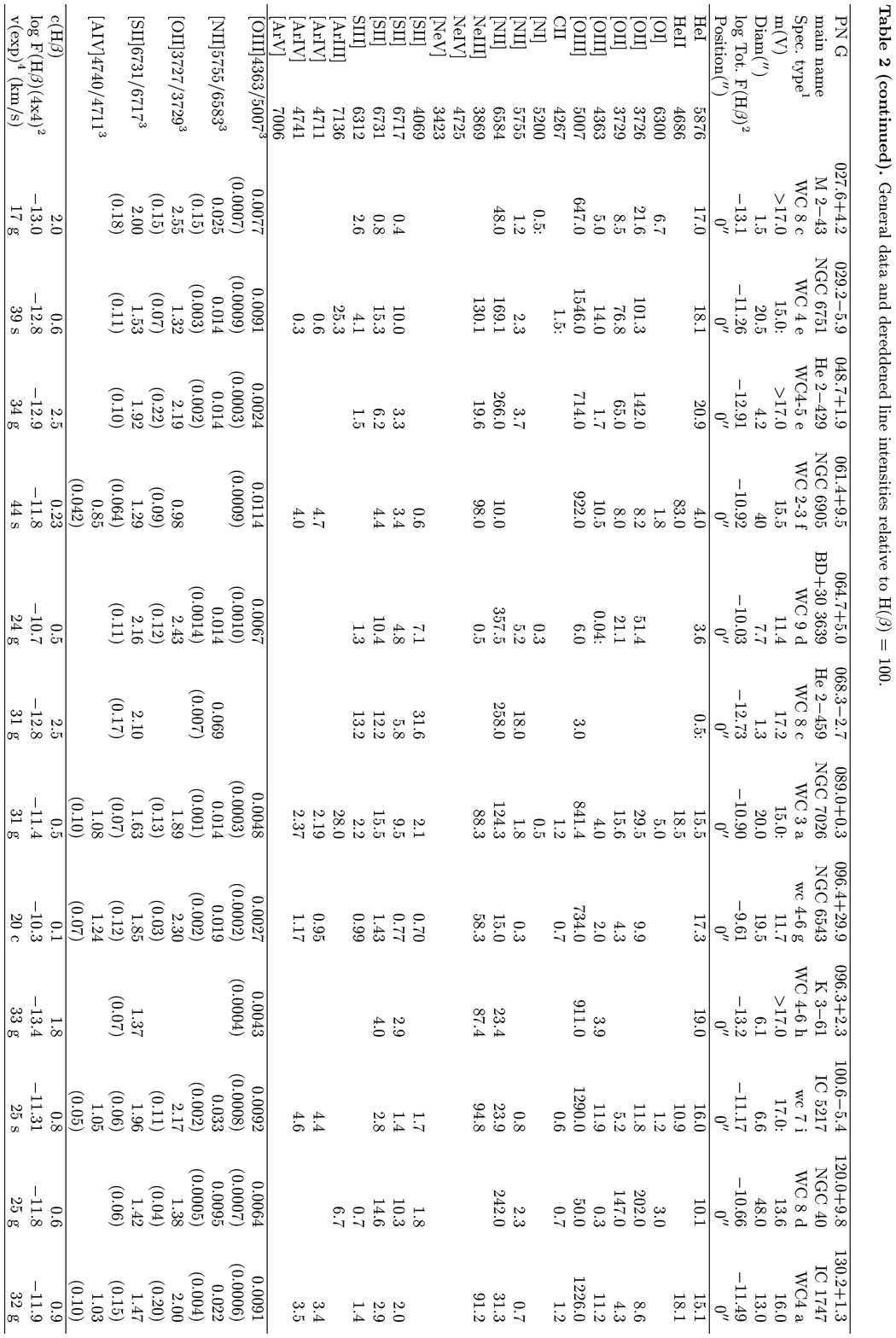}}
\end{figure*}

\begin{figure*}
\resizebox{19.5cm}{!}{\includegraphics{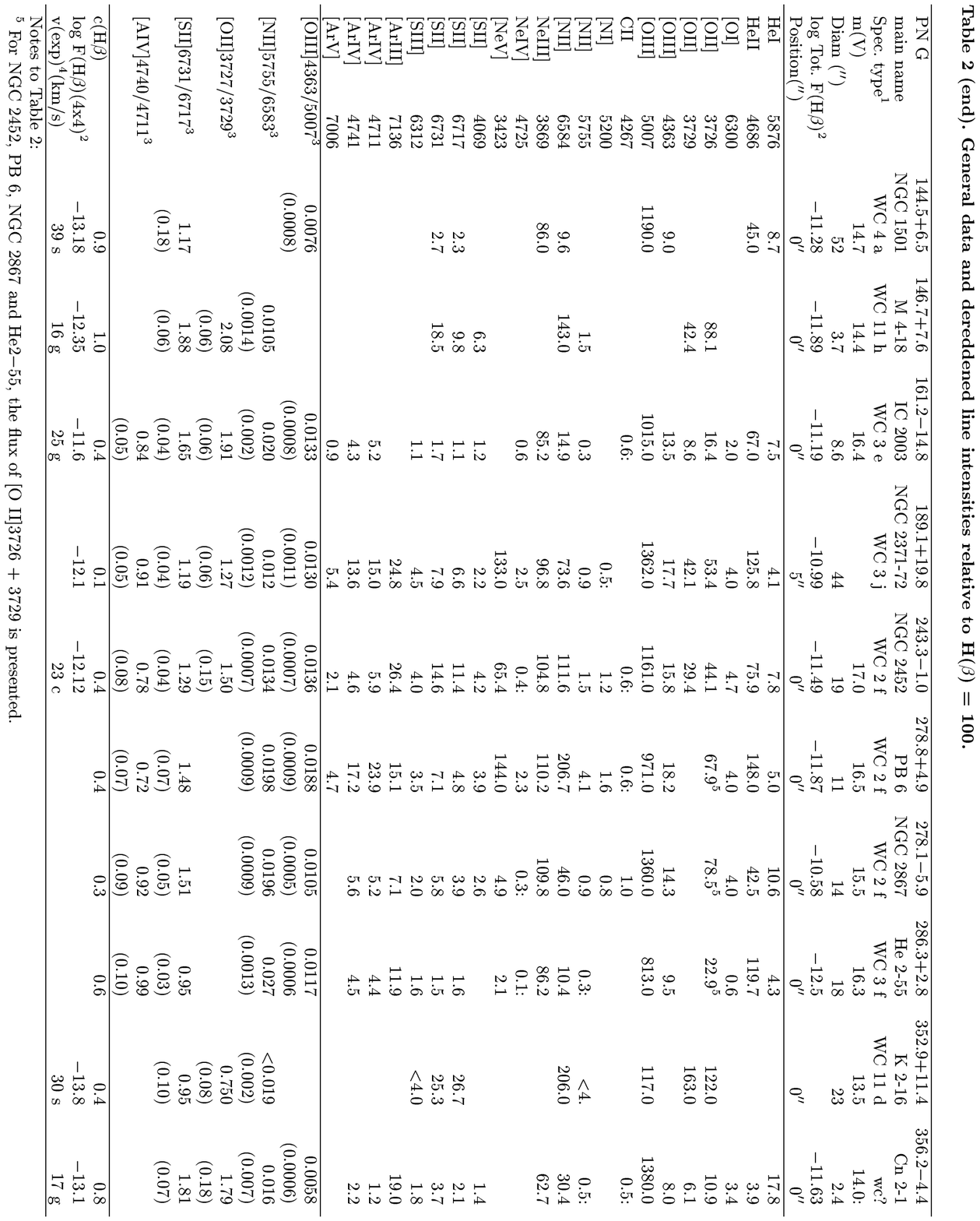}}
\end{figure*}

\onecolumn
\begin{table}
{{\bf Table 3}. Physical Conditions and Abundances.}
\begin{flushleft}
\begin{tabular}{lrrrrrrrrr}
\hline
		
name &	SwSt 1	 &       NGC 6369&	Cn 1-5&	         Hb 4&	         M 1-25&        M 3-15&	        NGC 6629&	NGC 6578&	M 1-32\\				 PN G &	1.5 $-$6.7&	2.4 +5.8&	2.2 $-$9.4&	3.1 +2.9&	4.9 +4.9&	6.8 +4.1&	9.4 $-$5.0&	10.8 $-$1.8&	11.9 +4.2\\				  &&&\\
n[SII]&	2.80E+05&	3.22E+03&	4.70E+03&	4.60E+03&	1.04E+04&		&       3.47E+03&	2.27E+03&	6.72E+03\\
&	2.33E+04&	2.64E+03&	4.10E+03&	3.66E+03&	6.60E+03&		&       2.44E+03&	1.87E+03&	5.67E+03\\	 
&	3.00E+05&	3.95E+03&	5.41E+03&	5.90E+03&	1.98E+04&		&       5.06E+03&	2.77E+03&	8.09E+03\\		 
&&&\\
n[OII]&	8.09E+03&	1.73E+03&	2.01E+03&	1.62E+03&	3.88E+03&	6.33E+03&	1.10E+03&	8.83E+02&	2.25E+03\\	 
	&6.21E+03&	1.39E+03&	1.52E+03&	1.03E+03&	2.51E+03&	2.44E+03&	7.86E+02&	6.60E+02&	1.42E+03\\	 
	&1.09E+04&	2.13E+03&	2.61E+03&	2.40E+03&	6.17E+03&	2.94E+04&	1.47E+03&	1.13E+03&	3.44E+03\\				 &&&&\\
n[ArIV]&	&	5.22E+03&		&	5.86E+03&		&		&		&	3.31E+03&	 \\
	&	&	3.82E+03&		&	4.40E+03&		&		&		&	2.25E+03&	 \\
	&	&	6.67E+03&		&	7.37E+03&		&		&		&	4.41E+03&	  \\
&&\\
T[OIII]&1.06E+04&	1.01E+04&	1.09E+04&	9.62E+03&	7.97E+03&	8.75E+03&	8.38E+03&	7.86E+03&	9.99E+03\\ 
&	1.01E+04&	9.91E+03&	1.06E+04&	9.28E+03&	7.75E+03&	8.39E+03&	8.13E+03&	7.75E+03&	9.68E+03 \\
&	1.10E+04&	1.03E+04&	1.13E+04&	9.95E+03&	8.18E+03&	9.09E+03&	8.62E+03&	7.98E+03&	1.03E+04\\
&&&&\\ 
T[NII]&	1.15E+04&	1.20E+04&	9.07E+03&	1.17E+04&	8.34E+03&	1.18E+04&		&	1.42E+04&	9.73E+03 \\
&	1.09E+04&	1.14E+04&	8.72E+03&	1.12E+04&	8.05E+03&	9.02E+03&		&	1.34E+04&	9.33E+03\\ 
&	1.20E+04&	1.26E+04&	9.40E+03&	1.22E+04&	8.63E+03&	1.26E+04&		&	1.51E+04&	1.01E+04 \\
&&&\\  
O++/O+&	3.11E$-$01&	1.57E+01&	8.28E$-$01&	3.39E+01&	4.36E+00&	3.88E+01&	1.35E+01&	2.40E+02&	1.21E+00\\ 
&	2.36E$-$01&	1.36E+01&	7.10E$-$01&	2.96E+01&	3.17E+00&	6.54E+00&	1.28E+01&	2.07E+02&	1.01E+00 \\
&	3.81E$-$01&	1.79E+01&	9.51E$-$01&	3.83E+01&	5.41E+00&	6.39E+01&	1.42E+01&	2.74E+02&	1.41E+00 \\
&&&&\\  
He/H&	3.99E$-$02&	1.14E$-$01&	1.16E$-$01&	1.38E$-$01&	1.35E$-$01&	1.40E$-$01&	9.34E$-$02&	1.56E$-$01&	1.26E$-$01\\
&&&&\\ 
O/H&	3.19E$-$05&	4.71E$-$04&	4.29E$-$04&	5.28E$-$04&	4.93E$-$04&	5.29E$-$04&	4.16E$-$04&	6.57E$-$04&	2.15E$-$04\\ 
&	2.72E$-$05&	4.45E$-$04&	3.99E$-$04&	4.71E$-$04&	4.52E$-$04&	4.61E$-$04&	3.72E$-$04&	6.19E$-$04&	2.01E$-$04\\ 
&	4.02E$-$05&	5.01E$-$04&	4.68E$-$04&	5.99E$-$04&	5.44E$-$04&	6.74E$-$04&	4.71E$-$04&	6.99E$-$04&	2.36E$-$04\\ 
&&&&\\ 
N/O&	8.03E$-$01&	4.72E$-$01&	3.17E$-$01&	1.23E+00&	8.11E$-$01&	5.97E$-$01&	8.84E$-$02&	6.58E$-$01&	1.09E+00 \\
&	6.63E$-$01&	4.39E$-$01&	2.94E$-$01&	1.11E+00&	6.46E$-$01&	1.82E$-$01&	8.31E$-$02&	6.10E$-$01&	9.49E$-$01\\ 
&	9.31E$-$01&	5.04E$-$01&	3.39E$-$01&	1.34E+00&	9.51E$-$01&	8.72E$-$01&	9.34E$-$02&	7.05E$-$01&	1.22E+00\\ 
& &&\\
Ne/O&		&	2.03E$-$01&	4.01E$-$01&	1.60E$-$01&	3.43E$-$02&	1.24E$-$01&	1.33E$-$01&	1.96E$-$01&	3.71E$-$02\\ 
&		&	2.00E$-$01&	3.94E$-$01&	1.56E$-$01&	3.36E$-$02&	1.21E$-$01&	1.30E$-$01&	1.94E$-$01&	3.64E$-$02\\ 
&		&	2.05E$-$01&	4.10E$-$01&	1.64E$-$01&	3.52E$-$02&	1.28E$-$01&	1.37E$-$01&	1.99E$-$01&	3.79E$-$02\\ 
\hline 
\end{tabular}
\end{flushleft}
\end{table}

\begin{table}
{{\bf Table 3 (continued)}. Physical Conditions and Abundances.}
\begin{flushleft}
\begin{tabular}{lrrrrrrrrr}
\hline		 

name&	NGC 6567&	PM 1-188&	M 3-30&		M 2-43&		NGC 6751&	He 2-429&	NGC 6905&	BD+30 3639&	He 2-459\\	 
PN G&	11.7 $-$0.6&	12.2 +4.9&	17.9 $-$4.8&	27.6 +4.2&   	29.2 $-$5.9&   	48.7 +1.9&	61.4 $-$9.5&	64.7 +5.0&	68.3 $-$2.7\\	 
&&\\ 
n[SII]&	4.36E+03&	5.01E+01&	5.24E+02&	1.01E+04&	2.73E+03&	6.81E+03&	1.53E+03&	2.12E+04&	1.74E+04\\	 
&	3.48E+03&	1.00E+01&	4.00E+02&	5.62E+03&	2.10E+03&	5.04E+03&	1.29E+03&	1.21E+04&	1.02E+04\\	 
&	5.54E+03&	2.35E+02&	6.59E+02&	2.52E+04&	3.55E+03&	9.72E+03&	1.81E+03&	6.14E+04&	6.19E+04\\	 
&&&\\ 
n[OII]&	4.97E+03&		&	1.01E+03&	7.39E+03&	1.06E+03&	3.64E+03&	4.51E+02&	5.99E+03\\				 
&	3.05E+03&		&	8.37E+02&	4.05E+03&	8.82E+02&	1.78E+03&	3.11E+02&	3.83E+03\\				 
&	8.60E+03&		&	1.19E+03&	1.63E+04&	1.25E+03&	8.03E+03&	6.01E+02&	1.01E+04	\\			  &&\\
n[ArIV]&4.72E+03&		&	1.99E+03&		&		&		&	1.77E+03&		 \\
&	3.67E+03&		&	4.91E+02&		&		&		&	7.35E+02&		 \\
&	5.80E+03&		&	3.53E+03&		&		&		&	2.82E+03&		 \\
&&\\ 								 
T[OIII]&9.49E+03&  		&	1.03E+04&	1.05E+04&	1.12E+04&	7.67E+03&	1.21E+04&	1.01E+04\\	 
&	9.35E+03&		&	1.01E+04&	1.02E+04&	1.08E+04&	7.48E+03&	1.17E+04&	9.61E+03&	 \\
&	9.63E+03&		&	1.06E+04&	1.08E+04&	1.15E+04&	7.85E+03&	1.25E+04&	1.05E+04&	\\
 &&\\								 
T[NII]&	9.38E+03&		&	1.16E+04&	1.17E+04&	9.93E+03&	9.49E+03&		&	9.19E+03&	2.00E+04\\
&	8.92E+03&		&	1.10E+04&	1.04E+04&	9.09E+03&	8.92E+03&		&	8.71E+03&	1.20E+04 \\
&	9.76E+03&		&	1.21E+04&	1.24E+04&	1.07E+04&	9.99E+03&		&	9.57E+03&	2.31E+04 \\
&&\\ 								 
O++/O+&	2.83E+01&		&	2.72E+01&	1.43E+01&	5.16E+00&	3.58E+00&	5.86E+01&	2.94E$-$02&\\	 
&	1.83E+01&		&	2.34E+01&	6.41E+00&	3.73E+00&	2.05E+00&	5.71E+01&	1.85E$-$02&\\	 
&	3.72E+01&		&	3.12E+01&	2.11E+01&	6.70E+00&	4.81E+00&	6.03E+01&	3.92E$-$02&\\	 
 &&\\								 
He/H&	8.79E$-$02&		&	1.71E$-$01&	1.10E$-$01&	1.28E$-$01&	1.40E$-$01&	1.10E$-$01&	2.37E$-$02&	2.02E$-$03\\ 
&&\\
O/H&	4.24E$-$04&		&	2.82E$-$04&	2.01E$-$04&	4.41E$-$04&	8.21E$-$04&	4.23E$-$04&	7.02E$-$05&	1.67E$-$07 \\
&	4.04E$-$04&		&	2.61E$-$04&	1.84E$-$04&	4.04E$-$04&	7.58E$-$04&	3.83E$-$04&	5.25E$-$05&	1.24E$-$07 \\
&	4.46E$-$04&		&	3.06E$-$04&	2.27E$-$04&	4.86E$-$04&	9.76E$-$04&	4.73E$-$04&	1.12E$-$04&	6.25E$-$07 \\
&&\\									 
N/O&	1.93E$-$01&		&	6.85E$-$01&	5.61E$-$01&	4.19E$-$01&	4.07E$-$01&	4.07E$-$01&	1.25E+00\\	 
&	1.42E$-$01&		&	6.36E$-$01&	3.22E$-$01&	3.58E$-$01&	2.73E$-$01&	3.86E$-$01&	9.04E$-$01\\	 
&	2.36E$-$01&		&	7.34E$-$01&	7.50E$-$01&	4.77E$-$01&	5.09E$-$01&	4.26E$-$01&	1.54E+00\\
&&\\									 
Ne/O&	1.27E$-$01&		&	2.63E$-$01&		&	2.26E$-$01&	2.00E$-$01&	2.72E$-$01&	2.39E$-$01\\	 
&	1.26E$-$01&		&	2.59E$-$01&		&	2.22E$-$01&	1.95E$-$01&	2.68E$-$01&	2.32E$-$01	 \\
&	1.28E$-$01&		&	2.68E$-$01&		&	2.31E$-$01&	2.04E$-$01&	2.78E$-$01&	2.47E$-$01	 \\
\hline 
\end{tabular}
\end{flushleft}
\end{table}

 \begin{table}
{{\bf Table 3 (continued)}. Physical Conditions and Abundances.}
\begin{flushleft}
\begin{tabular}{lrrrrrrrrr}
\hline		 			 
 
name&	NGC 7026&	NGC 6543&	K 3-61	&	IC 5217&	NGC 40&		IC 1747&	NGC 1501&	M 4-18&		IC 2003	\\		 
PN G&	89.0 +0.3&	96.4+29.9&	96.3 +2.3&	100.6 $-$5.4&	120.0 +9.8 & 	130.2 +1.3&  	144.5 +6.5&	146.7 +7.6&	161.2$-$14.8\\	 
&&\\ 
n[SII]&	3.25E+03&	5.64E+03&	1.69E+03&	8.85E+03&	1.98E+03&	2.38E+03&	1.02E+03&	6.35E+03&	3.84E+03\\ 
&	2.76E+03&	4.05E+03&	1.41E+03&	7.41E+03&	1.70E+03&	1.65E+03&	5.58E+02&	5.36E+03&	3.40E+03 \\
&	3.83E+03&	8.32E+03&	2.01E+03&	1.08E+04&	2.29E+03&	3.41E+03&	1.64E+03&	7.64E+03&	4.35E+03 \\
&&\\ 
n[OII]&	2.55E+03&	4.42E+03&		&	4.15E+03&	1.12E+03&	3.23E+03&		&	3.27E+03&	2.99E+03\\
&	1.73E+03&	3.98E+03&		&	2.91E+03&	1.01E+03&	1.76E+03&		&	2.71E+03&	2.27E+03 \\
&	3.70E+03&	4.93E+03&		&	6.02E+03&	1.24E+03&	5.89E+03&		&	3.95E+03&	3.91E+03 \\
&&\\ 
n[ArIV]&4.14E+03&	5.52E+03&		&	4.08E+03&		&	3.88E+03&		&		&	1.64E+03 \\
&	3.04E+03&	4.71E+03&		&	3.44E+03&		&	2.66E+03&		&		&	4.03E+02 \\
&	5.27E+03&	6.35E+03&		&	4.74E+03&		&	5.14E+03&		&		&	2.91E+03 \\
 &&\\
T[OIII]&9.16E+03&	7.90E+03&	8.91E+03&	1.12E+04&	1.00E+04&	1.12E+04&	1.05E+04&		&	1.28E+04\\ 
&	9.00E+03&	7.83E+03&	8.66E+03&	1.09E+04&	9.66E+03&	1.09E+04&	1.02E+04&		&	1.25E+04\\ 
&	9.31E+03&	7.98E+03&	9.14E+03&	1.15E+04&	1.03E+04&	1.14E+04&	1.09E+04&		&	1.31E+04 \\
&&\\ 
T[NII]&	9.81E+03&	1.09E+04&		&	1.44E+04&	8.62E+03&	1.19E+04&	&		8.63E+03&	1.12E+04\\ 
&	9.53E+03&	1.05E+04&		&	1.38E+04&	8.46E+03&	1.08E+04&		&	8.23E+03&	1.06E+04 \\
&	1.01E+04&	1.12E+04&		&	1.51E+04&	8.77E+03&	1.29E+04&		&	9.02E+03&	1.17E+04\\ 
 &&\\ 
O++/O+&	1.47E+01&	7.54E+01&		&	9.39E+01&	6.42E$-$02&	7.86E+01&		&		&	2.14E+01 \\
&	1.23E+01&	6.76E+01&		&	7.56E+01&	5.91E$-$02&	5.60E+01&		&		&	1.84E+01 \\
&	1.68E+01&	8.35E+01&		&	1.10E+02&	7.05E$-$02&	9.92E+01&		&		&	2.45E+01 \\
 &&\\ 
He/H&	1.24E$-$01&	1.16E$-$01&	1.31E$-$01&	1.15E$-$01&	7.11E$-$02&	1.18E$-$01&	1.05E$-$01&		&	1.13E$-$01\\ 
&&\\ 
O/H&	4.67E$-$04&	6.17E$-$04&	4.73E$-$04&	3.30E$-$04&	2.84E$-$04&	3.29E$-$04&	4.91E$-$04&	1.44E$-$04&	2.99E$-$04 \\
&	4.41E$-$04&	5.94E$-$04&	4.29E$-$04&	3.01E$-$04&	2.66E$-$04&	3.07E$-$04&	4.44E$-$04&	1.17E$-$04&	2.82E$-$04 \\
&	4.96E$-$04&	6.43E$-$04&	5.27E$-$04&	3.63E$-$04&	3.05E$-$04&	3.55E$-$04&	5.48E$-$04&	1.82E$-$04&	3.18E$-$04\\ 
&&\\ 
N/O&	1.01E+00&	3.96E$-$01&		&	7.67E$-$01&	2.33E$-$01&	1.15E+00&		&	2.98E$-$01&	2.61E$-$01\\ 
&	8.90E$-$01&	3.75E$-$01&		&	6.50E$-$01&	2.25E$-$01&	8.92E$-$01&		&	2.71E$-$01&	2.37E$-$01\\ 
&	1.13E+00&	4.17E$-$01&		&	8.71E$-$01&	2.42E$-$01&	1.37E+00&		&	3.23E$-$01&	2.84E$-$01\\ 
&&\\ 
Ne/O&	3.22E$-$01&	2.74E$-$01&	3.01E$-$01&	1.98E$-$01&		&	2.00E$-$01&	2.02E$-$01&		&	2.10E$-$01\\ 
&	3.19E$-$01&	2.72E$-$01&	2.95E$-$01&	1.94E$-$01&		&	1.98E$-$01&	1.98E$-$01&		&	2.08E$-$01\\ 
&	3.27E$-$01&	2.77E$-$01&	3.07E$-$01&	2.01E$-$01&		&	2.03E$-$01&	2.06E$-$01&		&	2.13E$-$01\\
\hline 
\end{tabular}
\end{flushleft}
\end{table}
 
 \begin{table}
{{\bf Table 3 (end)}. Physical Conditions and Abundances.}
\begin{flushleft}
\begin{tabular}{lrrrrrrrrr}
\hline		 			 
name&	NGC 2371-72&	NGC 2452&	PB 6&		NGC 2867&	He 2-55&	K 2-16&		Cn 2-1 \\
PN G&	189.1+19.8&	243.3 $-$1.0&	278.8 +4.9&	278.1 $-$5.9&	286.3 +2.8&	352.9+11.4&  	356.2 $-$4.4\\ 
&&\\ 
n[SII]&	1.19E+03&	1.59E+03&	2.69E+03&	2.66E+03&	5.16E+02&	5.04E+02&	5.32E+03\\ 
&	1.07E+03&	1.43E+03&	2.27E+03&	2.39E+03&	4.50E+02&	3.00E+02&	4.39E+03\\
&	1.31E+03&	1.77E+03&	3.18E+03&	2.97E+03&	5.85E+02&	7.37E+02&	6.54E+03\\ 
&&\\ 
n[OII]&	1.00E+03&	1.57E+03&		&		&		&	1.03E+02&	2.24E+03\\ 
&	8.34E+02&	1.02E+03&		&		&		&	2.91E+01&	1.33E+03\\ 
&	1.18E+03&	2.26E+03&		&		&		&	1.81E+02&	3.61E+03\\ 
&&\\ 
n[ArIV]&2.52E+03&	9.34E+02&	1.82E+02&	2.60E+03&	3.53E+03&		\\ 
&	1.95E+03&	1.00E+01&	1.00E+01&	1.48E+03&	2.27E+03&		 \\
&	3.09E+03&	1.91E+03&	1.12E+03&	3.75E+03&	4.82E+03&		  \\
 &&\\
T[OIII]&1.27E+04&	1.30E+04&	1.48E+04&	1.17E+04&	1.22E+04&		&	9.70E+03\\ 
&	1.23E+04&	1.27E+04&	1.45E+04&	9.45E+03&	1.20E+04&		&	9.40E+03\\ 
&	1.31E+04&	1.32E+04&	1.51E+04&	1.37E+04&	1.25E+04&		&	9.99E+03 \\
&&\\ 
T[NII]&	9.41E+03&	9.74E+03&	1.13E+04&	1.12E+04&	1.38E+04&	1.17E+04&	1.06E+04\\ 
&	9.04E+03&	9.54E+03&	1.11E+04&	1.10E+04&	1.34E+04&	1.17E+04&	1.04E+04\\ 
&	9.77E+03&	9.95E+03&	1.16E+04&	1.15E+04&	1.42E+04&	1.18E+04&	1.08E+04\\ 
 &&\\ 
O++/O+&	5.37E+00&	6.01E+00&	6.24E+00&	1.15E+01&	4.95E+01&	4.45E$-$01&	7.42E+01\\ 
&	4.62E+00&	5.33E+00&	5.81E+00&	8.30E+00&	4.63E+01&	4.36E$-$01&	6.04E+01\\ 
&	6.15E+00&	6.68E+00&	6.66E+00&	2.00E+01&	5.27E+01&	4.54E$-$01&	8.68E+01\\ 
 &&\\ 
He/H&	1.51E$-$01&	1.27E$-$01&	1.75E$-$01&	1.12E$-$01&	1.48E$-$01&		&	1.05E$-$01\\ 
&&\\ 
O/H&	7.91E$-$04&	3.75E$-$04&	3.89E$-$04&	4.12E$-$04&	4.24E$-$04&	7.89E$-$05&	5.57E$-$04\\ 
&	7.36E$-$04&	3.57E$-$04&	3.73E$-$04&	2.78E$-$04&	4.03E$-$04&	7.77E$-$05&	5.04E$-$04\\ 
&	8.55E$-$04&	3.94E$-$04&	4.08E$-$04&	8.04E$-$04&	4.48E$-$04&	8.02E$-$05&	6.23E$-$04\\ 
&&\\ 
N/O&	3.12E$-$01&	6.07E$-$01&	1.40E+00&	2.67E$-$01&	3.50E$-$01&	4.86E$-$01&	7.77E$-$01\\ 
&	2.90E$-$01&	5.53E$-$01&	1.32E+00&	2.58E$-$01&	3.38E$-$01&	4.77E$-$01&	6.64E$-$01\\ 
&	3.33E$-$01&	6.59E$-$01&	1.47E+00&	2.76E$-$01&	3.62E$-$01&	4.96E$-$01&	8.78E$-$01\\ 
&&\\ 
Ne/O&	1.78E$-$01&	2.25E$-$01&	2.68E$-$01&	2.12E$-$01&	2.71E$-$01&		&	1.34E$-$01\\ 
&	1.76E$-$01&	2.23E$-$01&	2.65E$-$01&	1.96E$-$01&	2.68E$-$01&		&	1.32E$-$01\\ 
&	1.81E$-$01&	2.27E$-$01&	2.70E$-$01&	2.43E$-$01&	2.73E$-$01&		&	1.37E$-$01\\ 
\hline 
\end{tabular}
\end{flushleft}
\end{table}

\twocolumn

{\bf Appendix: Notes on individual objects}

\subsection{K\,2-16}
K\,2-16, ionized by a [WC\,11] type star is a remarkable object.
The central star is  very bright with a visual magnitude of about 13 mag
(probably variable).
It was studied in detail by Leuenhagen et al. (1996) who derived a
temperature T$_*$ =
30,000~K, a terminal wind velocity of 300~km~s$^{-1}$, a mass loss rate of
4.4$\times 10^{-7}$ M$_\odot$ yr$^{-1}$
  and a chemical composition (in mass fraction) $\beta_{\rm H} \leq$ 1\%,
  $\beta_{\rm He}$= 45\%, 
$\beta_{\rm C}$= 50\% and $\beta_{\rm O}$=5\%. That is, the star is showing
helium burning products in
the atmosphere. According to these authors, the star
is very similar to other [WC 11] stars such as the nuclei of M\,4-18, He\,2-113
and others,
with a similar evolutionary status.

On the other hand, the nebula in K\,2-16 is quite different
showing a very faint and  extended shell with an angular diameter of about
20 arc sec. It presents a much lower density than nebulae around most 
of [WC]-late stars
  which are usually compact and very dense, and the shell shows a 
large expansion velocity.

  In our  spectra from the central zone, the star dominates the 
emission and its numerous and
intense lines are blended with the faint nebular lines, hiding them.
Therefore we extracted the nebular spectrum from the zone immediately 
outside the
stellar continuum.

The oxygen abundance determined for this object is only a lower limit 
due to the
faintness of [\ion{N}{ii}] 5755 for which only an upper limit is  given.

\subsection{He\,2-459}

This nebula, which has a [WC 8] central star, 
 was one of those which were considered by Acker et al. (1996) 
as candidates for the ``born again'' scenario, since its [\ion{S}{ii}]
ratio indicated a density of 3,600 cm$^{3}$,
lower than in other nebulae surounding stars of similar [WC] type.
Our observations from different observing runs consistently indicate
a [\ion{S}{ii}] density of 17,000 cm$^{3}$. In this object, 
which is highly reddened, 
the [\ion{O}{ii}] ratio is not available. Interestingly, this object
shows an extremely weak \ion{He}{i} $\lambda 5876$ emission line (about 0.005
of H$\beta$).
The nebula is also of very low excitation, with a  [\ion{O}{iii}] $\lambda 5007$
intensity of 0.03 of  H$\beta$. This is quite unexpected for a nebula
excited by a [WC\,8] central star, for which  
Leuenhagen \& Hamann (1998) have derived an effective temperature of 77,000 K,  
from a stellar atmosphere analysis. The \ion{H}{i}, \ion{He}{i} 
and \ion{He}{ii} Zanstra temperatures derived by these authors 
for their models of the star are 47,000, 45,000 and 16,000 K respectively.
Could it be that, because of the very large mass loss rate in this star 
(10$^{-5}$ M$_\odot$ yr$^{-1}$  assuming log L/L$_\odot$ = 3.70,  according 
to Leuenhagen \& Hamann 1998) and the large helium abundance in the atmosphere, 
the radiation shortward of the 24.6 eV would be much more efficiently blocked
that predicted? In any case, He\,2-459 deserves a much more thorough theoretical
and observational study.
 
\subsection{M\,1-32 and M\,3-15}
High velocity material has been detected in these objects.  In both
cases the line profiles show a strong narrow component with an almost
unresolved expansion velocity lower than 20~km~s$^{-1}$ overimposed on
faint blue and red wide wings.  In the case of M\,1-32, the wings extend
to about $\pm$100~km~s$^{-1}$, while in M\,3-15 the wings spread from 
$-$80 km~s$^{-1}$
to +90~km~s$^{-1}$.  Only the HWHM of the narrow profiles have been used in Fig.
1-l and 2.

Another object where high velocity ejections have been detected is Hb\,4.
L\'opez et al.  (1996)  found collimated outflows, several arcsec away from the
main body of this nebula with radial velocities of $\pm$150~km~s$^{-1}$.
According to these authors, the nebula presents an expansion velocity
of 21.5~km~s$^{-1}$, which is identical (within uncertainties) to our value
from the HWHM profile.

The nature of the high velocity material in M\,1-32 and M\,3-15 could be
different, as the emission comes from an unresolved central zone.
Spatially resolved observations are required to elucidate where this
high velocity emission comes from, but a bipolar or multipolar
ejection cannot be excluded.

\begin{acknowledgements}
  This work received partial support from the CNRS-CONACyT/M\'exico
agreement (grant E130-983), CONACyT (grant 32594-E) and DGAPA/UNAM (grants IN-\-109696 and IN-100799), the
University Paris 7 and the Observatoire de Paris-Meudon. MP thanks
the DAEC and GS the UNAM for their hospitality.
\end{acknowledgements}

\end{document}